\documentclass[usenatbib]{mn2e}
\usepackage{aasmacros}
\usepackage{graphicx}
\usepackage{pdfcomment}
\usepackage{xcolor} 
\usepackage{amsmath}
\pdfcommentsetup{draft}

\newcommand{\Mbh}{M_{\rm BH}}
\newcommand{\Mgal}{M_{\rm gal}}
\newcommand{\Mbhdot}{\langle \dot{M}_{\rm BH} \rangle}
\newcommand{\Mgaldot}{\dot{M}_{\rm gal}}
\newcommand{\zfid}{$z = 2$}

\begin{document}

\title[Quasar Demographics]{Comparing Simple Quasar Demographics Models}

\author[Veale et. al.]{Melanie Veale$^1$, Martin White$^{1,2}$, and Charlie Conroy$^3$\\
$^1$Departments of Physics and Astronomy,
  University of California, Berkeley, CA 94720, USA \\
$^2$Physics Division, Lawrence Berkeley National Laboratory,
  Berkeley, CA 94720, USA \\
$^3$Department of Astronomy \& Astrophysics,
  University of California, Santa Cruz, CA, 95060, USA
}

\date{MNRAS, accepted 3 September 2014}

\maketitle
\begin{abstract}

This paper explores several simple model variations for the connections
among quasars, galaxies, and dark matter halos for redshifts
$1<z<6$. A key component of these models is that we enforce a
self-consistent black hole (BH) history by tracking both
BH mass and BH growth rate at all redshifts. We connect objects across
redshift with a simple constant-number-density procedure, and choose a
fiducial model with a relationship between BH and galaxy growth rates
that is linear and evolves in a simple way with redshift.  Within this
fiducial model, we find the quasar luminosity function (QLF) by
calculating an ``intrinsic'' luminosity based on either the BH mass or
BH growth rate, and then choosing a model of quasar variability with
either a lognormal or truncated power-law distribution of
instantaneous luminosities. This gives four model variations, which we
fit to the observed QLF at each redshift. With the best-fit models in
hand, we undertake a detailed comparison of the four fiducial models,
and explore changes to our fiducial model of the BH-galaxy
relationship. Each model variation can successfully fit the observed
QLF, the shape of which is generally set by the ``intrinsic''
luminosity at the faint end and by the scatter due to variability at
the bright end. We focus on accounting for the reasons that physically
different models can make such similar predictions, and on identifying
what observational data or physical arguments are most essential in
breaking the degeneracies among models.

\end{abstract}

\begin{keywords}
quasars: general, quasars: supermassive black holes,
  galaxies: evolution, galaxies: star formation, galaxies: high
  redshift
\end{keywords}


\section{Introduction}
\label{sec:introduction}

Quasars are an important component of modern astrophysics, from their
role as extremely luminous objects useful for high redshift surveys to
their apparent influence on galaxy formation \citep[see
  e.g. the recent reviews of][]{ale12, kor13}. Many phenomenological
models have arisen to desribe the connections among quasars, the black
holes (BHs) that power them, and their host galaxies and dark matter
halos \citep[e.g.][]{EfsRee88, Car90, WyiLoe02, WyiLoe03, HaiCioOst04,
  Mar06, Lid06, Croton09, She09, BooSch10}. Despite active research
and a wealth of observational data, a single picture of quasar
demographics has yet to emerge.

Two recent models in particular have explored simple connections
between quasar activity and host galaxy properties. In \citet{CW13},
hereafter abbreviated ``CW13'', the model began by assuming a linear
relationship between BH mass and host galaxy mass, and calculated
quasar luminosities by assuming a single, mass-independent duty cycle
and Eddington ratio with some lognormal scatter. In \citet{H14},
hereafter abbreviated ``H14'', a very similar model assumed a linear
relationship between average BH accretion rate and galaxy star
formtion rate, and then found the quasar luminosity function (QLF) by
assuming a truncated power-law distribution of
instantaneous accretion rates and a constant radiative
efficiency. Both models, despite different perspectives on the
BH-galaxy connection and different assumptions about quasar
variability, were successful in explaining the basic properties of the
observed QLF, along with other observed quasar properties.

This paper aims to connect these models in a self-consistent framework
that tracks both BH mass and BH growth across redshift. With both
BH masses and average BH accretion rates in hand, we can make a direct
comparison between model types. The ``model space'' we consider has,
effectively, three ``dimensions'': the choice of BH-galaxy
relationship (including redshift evolution), the choice of whether to
connect quasar activity to BH masses or average BH accretion rates,
and the choice of quasar variability model. To facilitate the
exploration of this model space, we will make use of simplifying
assumptions wherever possible, while being mindful of where such
simplifications may not apply. In particular, our assumptions about
redshift evolution begin to break down at low redshift, so we will
restrict ourselves to the redshift range $1<z<6$. In this range, there
is much data available from large-scale (wide area, high redshift)
optical surveys \citep[e.g.][]{Wol03, Ric06, Cro09, Wil10, Ike11,
  Mas12, McG13, Ros13}, which makes the QLF for optical
(type-I) quasars a useful observable to choose as the ``input'' for
setting the best-fit parameters of each model variation.

Section \ref{sec:model} of this paper describes our fiducial model and
variations in detail; section \ref{sec:fid} compares the success of
each fiducial model variation in fitting the observed QLF; section
\ref{sec:notfid} explores variations beyond our fiducial model; and
section \ref{sec:summary} summarizes the major conclusions and
implications of our results. Where necessary, we use a $\Lambda$CDM
cosmology with $\Omega_M=0.28$ and $\Omega_\Lambda=0.72$, and assume
$h=0.7$. Stellar masses assume a \citet{Cha03} initial mass
function. Unless specificed otherwise, the $\log$ of any quantity is
taken to be $\log_{10}$.


\begin{figure*}
\begin{center}
\resizebox{6.5in}{!}{\includegraphics{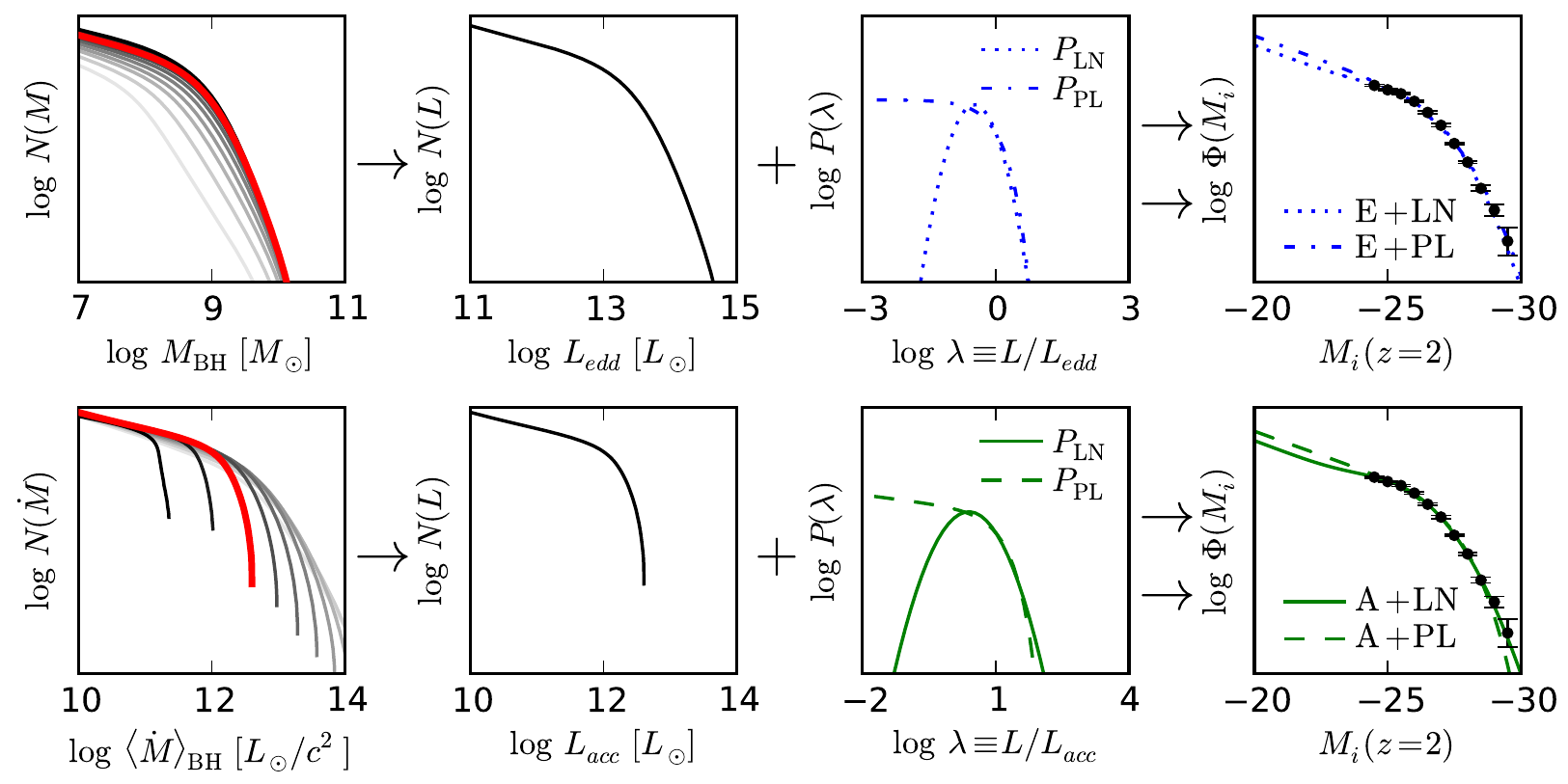}}
\end{center}
\caption{A schematic look at our model. The \textbf{left column} shows
  the BH properties (mass in the top panel, average growth rate in the
  bottom panel) for our fiducial ``growth-based evolution'' model of
  the BH-galaxy relationship. All of our observed redshifts are shown
  (see figure \ref{fig:multipanel} for the full list), with black at
  the lowest redshift ($z=1$) and lightest gray at the highest
  redshift ($z=6$). The remaining columns show only \zfid, which is
  highlighted in red in the first column. The \textbf{second column}
  shows the ``intrinsic'' QLF used in the ``Eddington model'' (top
  panel) and ``accretion model'' (bottom panel). These are simply the
  same curves from the left column, expressed in units of
  luminosity. The \textbf{third column} shows the distribution of
  observed luminosities (in units of ``intrinsic'' luminosity). The
  \textbf{right column} shows the results of convolving the
  ``intrinsic'' QLF in the second column with the distributions in the
  third column. This gives four model variations: \textbf{E+LN}, the
  dotted blue line, is the Eddington model convolved with
  $P_{\rm LN}$; \textbf{E+PL}, the dash-dotted blue line, is the Eddington
  model convolved with $P_{\rm PL}$; \textbf{A+LN}, the solid green
  line, is the accretion model convolved with $P_{\rm LN}$; \textbf{A+PL},
  the dashed green line, is the accretion model convolved with
  $P_{\rm PL}$. See table \ref{tab:terms} for a summary of the terminology.}
\label{fig:model}
\end{figure*}

\section{The model}
\label{sec:model}

\subsection{Galaxy mass and growth rate}
\label{sec:model_gal}

Our model begins with the halo mass functions (HMFs) from the fitting
functions of \citet{Tin08,Tin10}. These are translated into galaxy
stellar mass functions (SMFs) using the empirically constrained stellar
mass-halo mass relations from \citet{Beh12}. These relations are
calculated along with two components of scatter, an ``intrinsic''
scatter and ``observational'' scatter. For our purposes, we convolve
the SMF with only the ``intrinsic'' scatter, since we are not
interested in a direct comparison to the observed SMF.

These first steps are the same as the ones taken in CW13, but we add
the additional calculation of finding the mass growth rate of galaxies
across redshift. We will sometimes refer to this as the star formation
rate (SFR), although equating net mass growth with star formation is
only an approximation. The true SFR differs from the net mass growth
due to factors such as stellar mass loss and merging, which are
discussed in \citet{Beh12} but which we do not consider in detail in
this paper. To connect the SMFs across redshift, we use a matching
procedure that assumes the galaxy masses preserve rank order, and each
galaxy maintans a position in the SMF with constant number
density. With this assumption, we can obtain the galaxy growth rates
at each redshift using a simple central-difference approximation. At
very high masses, this yields a negative growth (which we set to zero
for the purposes of the model), indicating that our assumptions are
not accurate for such extreme objects. For low redshift, this negative
growth impacts more galaxies (all those above $10^{11} M_\odot$ for
$z=1$), so we restrict our analysis to the range $1 < z < 6$.

\subsection{Black hole-galaxy relations}
\label{sec:model_bh}

With a stellar mass history for each galaxy in hand, we must now
decide how to relate galaxies to their central BHs. To begin we
neglect scatter in the BH-galaxy relationship, which allows us to
directly apply our matching procedure across redshifts to the BHs as
well as the galaxies, without introducing additional complications.

Our fiducial model assumes a linear relationship between BH and galaxy
growth rates, similar to H14, but adds a scaling with redshift
similar to what was used in CW13.
\begin{align}
\Mbhdot &= 10^{-3.5} \Mgaldot (1+z)^2
\\ \implies \Mbhdot &= 10^{\alpha_G} \Mgaldot
\\ \text{where} \; \alpha_G(z) &\equiv -3.5 + 2 \log (1+z)
\end{align}
Where we implicitly assume that $\alpha_G$ is independent of mass for
this model. The choice of a local value of $\alpha_G=-3.5$ is
motivated by observations such as \citet{Raf11}, \citet{Mul12},
\citet{Che13}. The choice of redshift scaling is
discussed in more detail in section \ref{sec:ev}. We then integrate
this growth from redshift $z\sim8$ to obtain the BH masses, which gives us the
$\Mbh/\Mgal$ relationship and the BH mass function (BHMF) at each
redshift. However, the $\Mbh/\Mgal$ relationship is not a purely
linear one; it contains a mass dependence, which we fold into the
proportionality constant $\alpha_M$. To a very rough approximation,
ignoring both the mass dependence and additional redshift dependence,
$\alpha_M$ is similar to $\alpha_G$ with a small offset. (See the
appendix for a detailed discussion.)
\begin{align}
\Mbh &= 10^{\alpha_M} \Mgal
\\ \text{where} \; \alpha_M &= \alpha_M(z,\Mgal) 
\\ &\approx -3.2 + 2\log(1+z)
\end{align}
Since this model begins with a simple $\Mbhdot/\Mgaldot$ relationship
and requires integrating over redshift to find the BHMF, we refer to
it as the ``growth-based evolution'' model of BH-galaxy
relationships. Section \ref{sec:notfid} will discuss variations on
this fiducial model, including ``mass-based evolution'' and
``non-evolving'' models for the BH-galaxy relationship, which are also
discussed in the appendix.

The left column in figure \ref{fig:model} shows the BHMF and average growth
information for our fiducial model in the left two panels. Clear in
the bottom left panel is the effect of ``downsizing,'' meaning that
more massive objects ``complete'' their growth at earlier times. This
results in a dwindling supply of very quickly growing systems at low
redshift, and will be an important feature in our discussions in
section \ref{sec:notfid}.

\begin{table*}
\begin{center}
\fbox{\parbox{0.38\textwidth}{\textbf{BH-galaxy relationship} (see
    \ref{sec:model_bh} and appendix)\\
\rule[0.5ex]{\linewidth}{0.1ex}\\
``Growth-based evolution''
\begin{itemize}
\item Our fiducial model
\item Linear relationship between $\Mbhdot$ and $\Mgaldot$
\item The normalization ($\alpha_G$) evolves in a simple way with
  redshift
\item The relationship between $\Mbh$ and $\Mgal$ is derived by
  integrating across redshift
\end{itemize}
\rule[0.5ex]{\linewidth}{0.1ex} \\
``Mass-based evolution''
\begin{itemize}
\item Linear relationship between $\Mbh$ and $\Mgal$
\item The normalization ($\alpha_M$) evolves in a simple way with
  redshift
\item The relationship between $\Mbhdot$ and $\Mgaldot$ is derived
  by subtracting across redshifts
\end{itemize}
\rule[0.5ex]{\linewidth}{0.1ex}\\
``Non-evolving''
\begin{itemize}
\item Linear relationship between both $\Mbh/\Mgal$ and
  $\Mbhdot/\Mgaldot$
\item $\alpha_G$ and $\alpha_M$ are equal and independent of both
  redshift and mass
\end{itemize}
}}
$\bigotimes$
\fbox{\parbox{0.25\textwidth}{\textbf{Basis for ``intrinsic'' QLF}\\
(see \ref{sec:model_lum})\\
\rule[0.5ex]{\linewidth}{0.1ex}\\
``Eddington model''
\begin{itemize}
\item The ``intrinsic'' QLF is based on the BH mass via the Eddington
  luminosity.
\item Similar to CW13
\item Denoted by ``E'' in abbreviations
\end{itemize}
\rule[0.5ex]{\linewidth}{0.1ex}\\
``accretion model''
\begin{itemize}
\item The ``intrinsic'' QLF is based on the average BH growth rate via
  the total energy output available from the accreting mass.
\item Similar to H14
\item Denoted by ``A'' in abbreviations
\end{itemize}
}}
$\bigotimes$
\fbox{\parbox{0.27\textwidth}{\textbf{Luminosity distribution}\\
(see \ref{sec:model_lum})\\
\rule[0.5ex]{\linewidth}{0.1ex}\\
``Scattered lightbulb''
\begin{itemize}
\item Distribution $P_{\rm LN}(\lambda)$ is log-normal, with
  well-defined duty cycle and lifetime
\item Associated with step-function light curves
\item Similar to CW13
\item Denoted by ``LN'' in abbreviations
\end{itemize}
\rule[0.5ex]{\linewidth}{0.1ex}\\
``Luminosity-dependent lifetime''
\begin{itemize}
\item Distribution $P_{\rm PL}(\lambda)$ is a truncated power-law,
  with duty cycle and lifetime depending on choice of $\lambda_{min}$
\item Associated with more complex light curves than ``lightbulb'' models
\item Similar to H14
\item Denoted by ``PL'' in abbreviations
\end{itemize}
}}\\
\end{center}
\caption{A summary of terminology used to describe our model
  variations. Each complete model requires choosing one item
  from each column. There are thus four variations on the fiducial
  model (the \textbf{growth-based evolution} model), which we
  abbreviate as \textbf{E+LN}, \textbf{E+PL}, \textbf{A+LN},
  \textbf{A+PL} in figure legends where needed.}
\label{tab:terms}
\end{table*}

\subsection{Black hole luminosity distributions}
\label{sec:model_lum}

With the BH masses and growth rates in hand, we explore a total of
four simple options for obtaining the BH luminosities and thus the
QLF. First, we translate either the BH mass or growth rate into an
``intrinsic'' luminosity (based either on the Eddington luminosity or
the energy available from the accreting mass). We will refer to these as
the ``Eddington'' and ``accretion'' models, respectively. (They might
also be called ``mass-based'' and ``growth-based,'' but we wish to
avoid confusion with the different choices of BH-galaxy relationship
mentioned in section \ref{sec:model_bh}.) The conversions to
``intrinsic'' luminosity, called $L_{edd}$ for the Eddington model and
$L_{acc}$ for the accretion model, are defined as follows:
\begin{align}
\frac{L_{edd}}{L_\odot} &= 3.3 \times 10^4 \frac{\Mbh}{M_\odot} \\
\frac{L_{acc}}{L_\odot} &= \frac{\Mbhdot c^2}{L_\odot}
\end{align}
The ``intrinsic'' QLFs obtained from these conversions are illustrated
in the second column of figure \ref{fig:model}. This ``intrinsic'' QLF
is then convolved with a distribution of instantaneous luminosities to
capture the variable nature of quasars, and to encode parameters such
as the Eddington ratio, efficiency, and duty cycle. In the Eddington
model, this means defining a distribution of Eddington ratios,
i.e. $L/L_{edd}$. In the accretion model, this means defining a
distribution of a different ratio, $L/L_{acc}$.

For each choice of model for the ``intrinsic'' QLF, we compare two
distributions of instantaneous luminosity: a lognormal distribution
and a truncated power-law distribution. These
distributions encode information about quasar variability, and can
also be referred to as a ``scattered lightbulb'' model and a
``luminosity-dependent lifetime'' model, respectively, following the
terminology in \citet{Hop09}. The distributions are defined as follows:
\begin{align}
P_{\rm LN}(\lambda) &= \frac{A}{\sigma \sqrt{2\pi}} \exp \left(
\frac{-(\log \lambda - \log \lambda_0)^2}{2\sigma^2} \right) \label{eq:PLN} \\
P_{\rm PL}(\lambda) &= A \left( \frac{\lambda}{\lambda_0} \right)^{-\beta}
\exp \left(-\frac{\lambda}{\lambda_0} \right) \label{eq:PPL} \\
\text{where } \lambda \; &\text{is} \; \lambda_{edd} \equiv
\frac{L}{L_{edd}} \text{ or } \lambda_{acc} \equiv \frac{L}{L_{acc}}  \nonumber
\end{align}
We restrict $\beta$ to the range $0<\beta<1$, which covers the
possible distributions mentioned in H14. (We note that negative values
of $\beta$ give a $P_{\rm PL}$ distribution that is qualitatively
quite similar to $P_{\rm LN}$, so we do not consider them.)
These distributions are illustrated in the third column of figure
\ref{fig:model}. All parameters of the distribution ($A$, $\lambda_0$,
$\sigma$ or $\beta$) are assumed constant with $\Mbh$ (for the
Eddington model) or $\Mbhdot$ (for the accretion model), so that we
are convolving a single BH-independent distribution with the
``intrinsic'' QLF.  We tune these parameters separately at each
redshift to match the observed QLF, and this final QLF is illustrated
in the right column of figure \ref{fig:model}. These four combinations
of Eddington and accretion models with $P_{\rm LN}$ and $P_{\rm PL}$
distributions form the four variations of our fiducial model, which we
compare in detail in section \ref{sec:fid}.

The distribution parameters can be associated with physical
quantities: for example, in the Eddington models, $\lambda_{edd}$ is
the same as the Eddington ratio, while in the accretion models the
radiative efficiency is closely related to the average
$\lambda_{acc}$. We refer to \citet{Hop09} for a more detailed
discussion of the connection between $P(\lambda)$ and quasar
lifetimes, light curves, and triggering rates, but make use of the
terms for "lightbulb" models and "luminosity dependent lifetime
models." For the lognormal distributions, $A$ is simply related to the
duty cycle $f_{\rm on}$, and each accretion episode can be modeled as
a "lightbulb" (a step-function light curve) with luminosity drawn from
$P_{\rm LN}$, so we refer to these as scattered lightbulb models. For
the power-law distributions the duty cycle and quasar lifetime depend
on a choice of lower bound $\lambda_{min}$, and the light curve is not
a simple lightbulb model, so we refer to these as luminosity-dependent
lifetime models. We choose $\lambda_{min}=10^{-3}\lambda_0$ throughout
the paper. This value is a somewhat arbitrary choice, since any
$\lambda_{min}$ smaller than $10^{-3}\lambda_0$ can be chosen with no
effect on the QLF fit. (Larger choices of $\lambda_{min}$ begin to
have a small effect on the faint end of the QLF.) Very small values of
$\lambda_{min}$ can result in a duty cycle greater than one when
paired with very negative values of beta, but all of our best-fit
values fall within a reasonable range. As an example, for
$\lambda_{min}=10^{-5}\lambda_0$, the duty cycle becomes greater than
one for approximately $\beta<-0.4$.

The combination of the Eddington model with $P_{\rm LN}$ is very similar
to the fiducial model of CW13, whereas the accretion model with
$P_{\rm PL}$ is very similar to H14. However, in both cases we make
slightly different assumptions about the BH-galaxy relationship,
since our fiducial ``growth-based evolution'' model does not exactly
match either CW13 or H14.

\subsection{Model summary}
\label{sec:model_summary}

In summary, we have chosen a ``growth-based evolution'' model as our
fiducial model of the BH-galaxy connection, and defined four
variations on that model. The steps in each of the four variations are
illustrated in figure \ref{fig:model}, which uses the following
abbreviations: E+LN for the Eddington model with $P_{\rm LN}$
distribution, E+PL for the Eddington model with $P_{\rm PL}$ distribution,
A+LN for the accretion model with $P_{\rm LN}$ distribution, and A+PL for
the accretion model with $P_{\rm PL}$ distribution.

Each model variation has three free parameters associated with the
luminosity distribution $P(\lambda)$: $A$, $\lambda_0$, and
$\sigma$ (for $P_{\rm LN}$) or $\beta$ (for $P_{\rm PL}$). These parameters
are tuned at each redshift to match the observed QLF, and the
results are discussed in section \ref{sec:fid}.

We also explore beyond our fiducial model of the BH-galaxy connection
in section \ref{sec:notfid}, by considering ``mass-based evolution''
or ``non-evolving'' approaches, and discussing the impact of other
model assumptions such as neglecting scatter in the BH-galaxy
relationship.

A summary of the terminology used to identify the model variations is
shown in table \ref{tab:terms}.


\begin{figure*}
\begin{center}
\resizebox{6.5in}{!}{\includegraphics{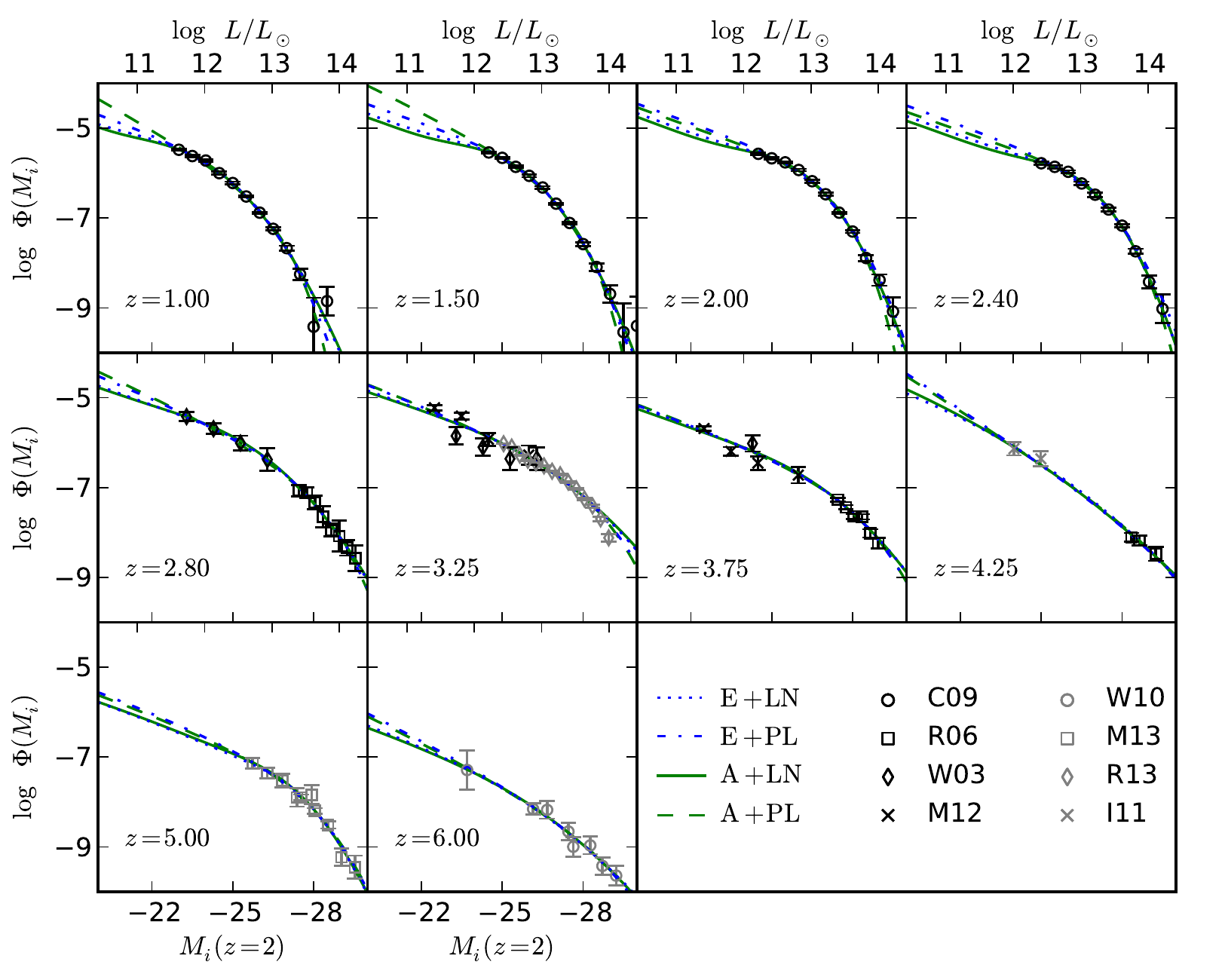}}
\end{center}
\caption{Fit results for each redshift. The lines representing each
  model variation follow the same abbreviations, colors, and line
  styles as in figure \ref{fig:model}. The data references are as
  follows: \textbf{C09} (black circle) is \citet{Cro09}, \textbf{R06}
  (black square) is \citet{Ric06}, \textbf{W03} (black diamond) is
  \citet{Wol03}, \textbf{M12} (black X) is \citet{Mas12}, \textbf{W10}
  (gray circle) is \citet{Wil10}, \textbf{M13} (gray square) is
  \citet{McG13}, \textbf{R13} (gray diamond) is \citet{Ros13},
  \textbf{I11} (gray X) is \citet{Ike11}.}
\label{fig:multipanel}
\end{figure*}

\section{Fiducial model variations}
\label{sec:fid}

\subsection{Fitting the QLF}
\label{sec:fit}

Figure \ref{fig:multipanel} shows the resulting best-fit QLF for each
model at each redshift, along with the compiled data. (The figure
caption lists references for each of the 8 sources of data.) To
compare our model to this data, we use the relation from
\cite{She++09} between $i$-band magnitude at $z=2$ and bolometric
luminosity, in terms of $L_\odot$:
\begin{equation}
M_i (z=2) = 6.04 - 2.5 \log \; (L/L_\odot)
\end{equation}
See e.g. the appendix of \citet{Ros13} for the filter transformations
and k-corrections necessary for expressing all of the data in terms of
$M_i(z=2).$

\begin{figure}
\begin{center}
\resizebox{3.5in}{!}{\includegraphics{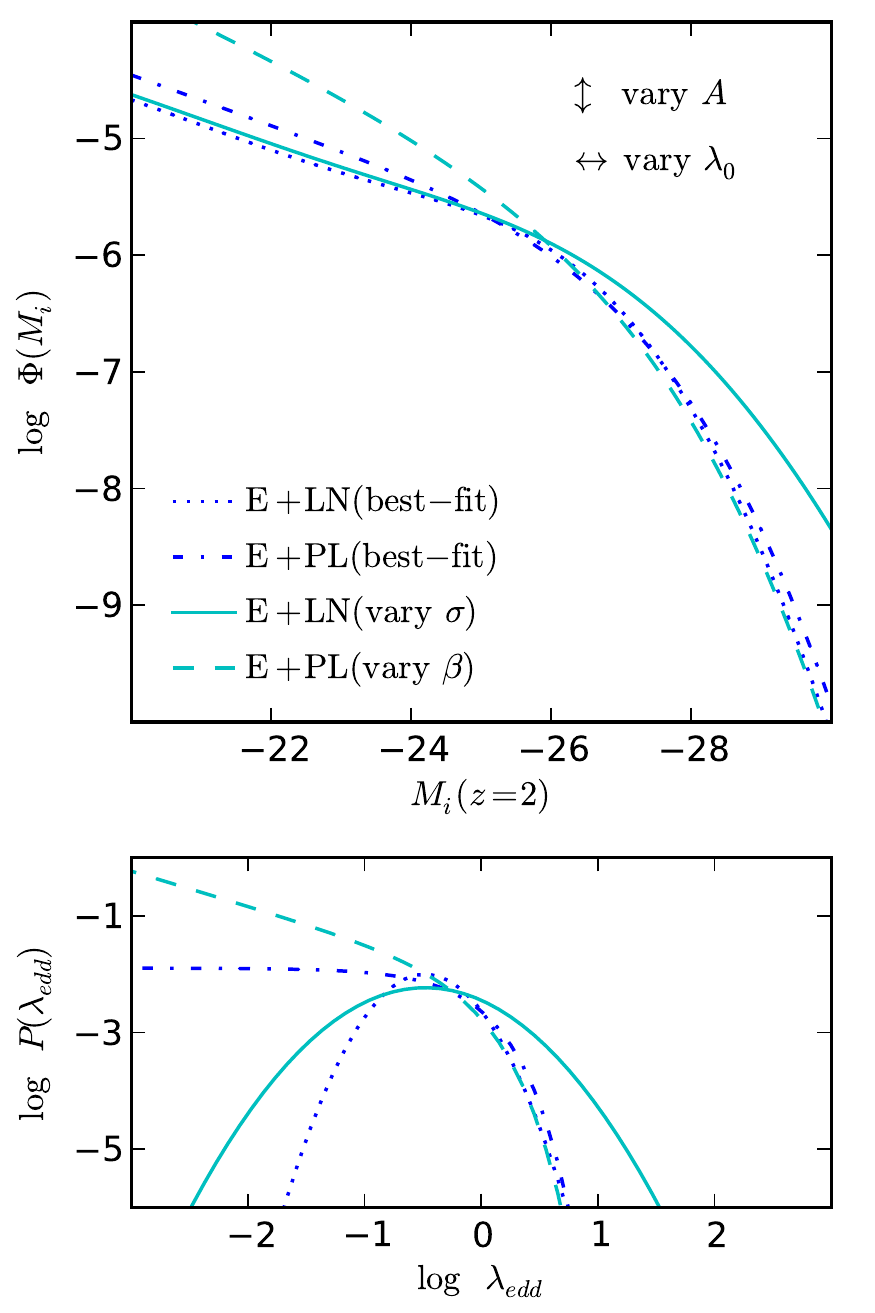}}
\end{center}
\caption{An example of how varying the luminosity distribution
  $P(\lambda)$ changes the overall QLF. See equations \ref{eq:PLN} and
  \ref{eq:PPL} for the parameterizations of $P_{\rm LN}$ (solid and
  dotted lines) and $P_{\rm PL}$ (dashed and dash-dotted lines). This
  shows the best-fit QLF (blue lines) of the Eddington model at \zfid,
  and the QLF given by varying $\sigma$ or $\beta$ (cyan lines). For
  both $P_{\rm LN}$ and $P_{\rm PL}$, varying $A$ simply moves the QLF
  up or down, and varying $\lambda_0$ moves it left or right.}
\label{fig:vary_params}
\end{figure}

It is immediately apparent that all four model variations fit the
observed QLF with similar levels of success. In order to disentangle
the reasons why four physically distinct models can make such similar
predictions, we will make a separate detailed analysis of the bright
and faint ends of the QLF. Figure \ref{fig:vary_params} will serve as
a useful reference point to these discussions, as it shows
qualitatively the substantial freedom our luminosity distributions
provide in fitting the observed QLF. The parameters $A$ and
$\lambda_0$ allow us to adjust the QLF horizontally and vertically,
while $\sigma$ and $\beta$ allow separate variation in the bright and
faint ends of the QLF. Figure \ref{fig:vary_params} uses the Eddington
model at \zfid $\;$as an example, but the same qualitative features
apply at all redshifts and for the accretion model as well.

\subsection{The bright end of the QLF}
\label{sec:bright}

We can look back at the second column of figure \ref{fig:model}, to see that
the ``intrinsic'' QLF falls off more quickly in the accretion model
than in the Eddington model, especially at low redshift. Physically,
this is due to the impact of ``downsizing'' on BH growth rates: while
the massive end of the BHMF remains relatively stable, there is a
dwindling supply of very quickly growing objects at low redshift, because
the more massive objects are already ``in place.'' However, in spite
of this difference in the ``intrinsic'' QLF, both the Eddington and
accretion models have similar success in fitting the QLF at the bright
end. This suggests that it is scatter, not the ``intrinsic'' QLF, that
sets the bright end of the observed QLF.

In figure \ref{fig:vary_params}, we can see that adjusting $\sigma$
can have a large effect on the bright end of the QLF, whereas
adjusting $\beta$ does not. However, the best-fit distributions, shown
in blue, are very similar at the large $\lambda$ end for both $P_{\rm LN}$ and
$P_{\rm PL}$. This shows that the bright end of the $P(\lambda)$ distribution is
well-constrained by the QLF, and that the amount of ``scatter'' in the
luminosity at the exponential cutoff of $P(\lambda)$ is, somewhat
coincidentally, approximately the right amount of scatter to fit the
QLF. This was illustrated in H14 as well, where most of the
discrepancy between the ``intrinsic'' QLF (in H14, based on the
observed galaxy star-formation rate) and the observed QLF could be
accounted for by the quasar variability in $P(\lambda)$.

The lack of flexibility in $P_{\rm PL}$ for adjusting the bright end of
the QLF does manifest as a slight under-prediction of the bright end
of the QLF at certain redshifts. However, our fiducial model neglects
scatter in the BH-galaxy relationship: including this additional
scatter would likely be enough to resolve this under-prediction in the
case of $P_{\rm PL}$, while in the case of $P_{\rm LN}$ it would be easy to
decrease $\sigma$ to compensate for increased scatter in the rest of
the model. In other words, the bright end of the $P(\lambda)$ is
well-constrained by the QLF in our model, but in general it is only
the combination of $P(\lambda)$ and scatter in the BH-galaxy
relationship that is actually well constrained.

The conclusion that ``the bright end of the QLF is set by scatter'' is
a general one, and regardless of the impact of scatter in the
BH-galaxy relationship, this fact allows two physically distinct
models to both successfully predict the observed QLF. Even though the
Eddington model and the accretion model look quite different at the
bright end of the ``intrinsic'' QLF, quasar variability erases this
difference in the observed QLF. The impact of scatter on the
properties of very luminous quasars also goes beyond the QLF; for
example, scatter can explain why hyperluminous quasars appear to live
in halo environments that are very similar to less luminous quasars,
as discussed in e.g. \citet{Tra12}, \citet{Fan13}. We
will return to the question of how scatter impacts the expected host
environments of luminous quasars in section \ref{sec:hosts}.

\subsection{The faint end of the QLF}
\label{sec:faint}

The characteristics of the faint end of the QLF are very different
from the bright end. Looking again at the second column of figure
\ref{fig:model}, we can see that the Eddington and accretion models
have very similar slopes at the faint end of the ``intrinsic'' QLF. In
contrast, the third column of figure \ref{fig:model} illustrates how
different the faint ends of the luminosity distributions are. The fact
that models using both $P_{\rm LN}$ and $P_{\rm PL}$ can successfully fit the
observed QLF suggests that the faint end of the QLF is set by the
``intrinsic'' QLF, not by $P(\lambda)$. In other words, the faint end
of $P(\lambda)$ is poorly constrained by the observed QLF, which is the
opposite of the situation at the bright end of the QLF.

However, the models using $P_{\rm PL}$ do diverge from the models using
$P_{\rm LN}$ at low redshift, below the range of the data. This occurs for
larger values of $\beta$ (closer to $\beta=1$), and suggests that the
faint end of the QLF may depend on quasar variability after all. Such
a situation is described in e.g. \citet{Hop09}, which contains an
extensive discussion of the differences between scattered
lightbulb models and luminosity dependent lifetime models.

There are two related reasons for the ambiguity in what governs the
slope of the faint end of the QLF. First, there are the limits to our
observational data. If we could obtain fainter data, e.g. down to
$M_i(z=2)=-20$ at $z=1$, we could better distinguish among
models. This is complicated, however, by the second problem: the
faint-end slope of the QLF may coincidentally be similar to
\emph{both} the slope of the BHMF \emph{and} to the slope of
$P(\lambda)$. (Because the BH growth rate is roughly proportional to
its mass, a fact we will return to in sections \ref{sec:zparams} and
\ref{sec:downsizing}, all statements about the BHMF in the Eddington
model apply to the accretion model as well.) The faint-end slopes of
our ``intrinsic'' QLFs correspond to roughly $\beta=0.4$ to $0.6$. The
slopes of $P_{\rm PL}$ suggested in H14 and \citet{Hop09} are also roughly
$\beta=0.4$ and $\beta=0.6$. This make it difficult to distinguish, at
the level of the observed QLF, between scattered lightbulb models and
luminosity-dependent lifetime models. Fairly precise measurements of
the faint-end slopes would be required to detect any difference. (A
``pure'' lightbulb model with no scatter at all, or a power-law
distribution with a hard cutoff at the bright end, might also be able
to reproduce the faint end of the QLF; however, they would fail at the
bright end, as we discussed in section \ref{sec:bright}.)

Another complication is the following: while a QLF slope that was
significantly different from the BHMF slope (thus suggesting that the
slope is governed by $P(\lambda)$) would be good evidence
\emph{for} a luminosity-dependent lifetime model, very similar slopes
are not necessarily evidence \emph{against} such a model. Similar
slopes in the BHMF and QLF only put constraints on how steep $\beta$
can be, because values near $\beta=0$ yield QLFs nearly
indistinguishable from those predicted by scattered lightbulb models.

Finally, it is important to mention that although we are discussing
the faint-end slope of the QLF as compared to the low-mass-end slope
of the BHMF, our fiducial model relies on a (roughly) linear
relationship between the BH and galaxy masses to calculate this
slope. In other words, we have implicitly assumed that the low-mass
slopes of the BHMF and SMF are the same. (Again, these statements
apply to the BH growth rates and SFR as well.) As a result, additional
QLF data at the faint end could rule out our fiducial lightbulb models
without ruling out \emph{all} scattered lightbulb models; a direct
measurement of the BHMF slope is required to truly make the
comparisons discussed above, and to attempt to rule out scattered
lightbulb models in general based on the QLF alone.

The result of all this ambiguity is that measurements of the QLF
do not necessarily constrain feeding models, quasar triggering
mechanisms, or other quasar physics that are encoded primarily in a
distribution like $P(\lambda)$.  We have shown that for our model, the
QLF can provide good constraints on the bright end of this
distribution, but not the faint end. In other words, faint quasars may
be \emph{either} high mass BHs at the low end of the $P(\lambda)$
distribution \emph{or} low mass BHs accreting at or near the Eddington
limit, as has been pointed out in other works \citep[e.g.][]{fan12,
  Ros13}.

There are several types of measurements useful for complementing the
QLF and constraining $P(\lambda)$. One choice is to measure
$P(\lambda)$ directly, as done in
e.g. \citet{kau09,Hop09,bon12,air12,air13,aza14}. Numerical
simulations, such as in \citet{nov11,nov12} and \citet{gab13}, can
also shed some light
on $P(\lambda)$. Another measurement choice is to
measure average host property as a function of luminosity, which we
discuss in detail in section \ref{sec:hosts}. Although we focus on
average galaxy SFR in section \ref{sec:hosts}, other host properties
can be used for a similar analysis such as host halo mass
\citep[e.g.][and references therein, or many other measurements of
  quasar clustering and bias]{she13}. Each of these approaches
generally suggests a ``luminosity-dependent lifetime'' model rather
than a ``scattered lightbulb'' model.

\subsection{Parameter correlations and trends}
\label{sec:zparams}

\begin{figure}
\begin{center} 
\resizebox{3.5in}{!}{\includegraphics{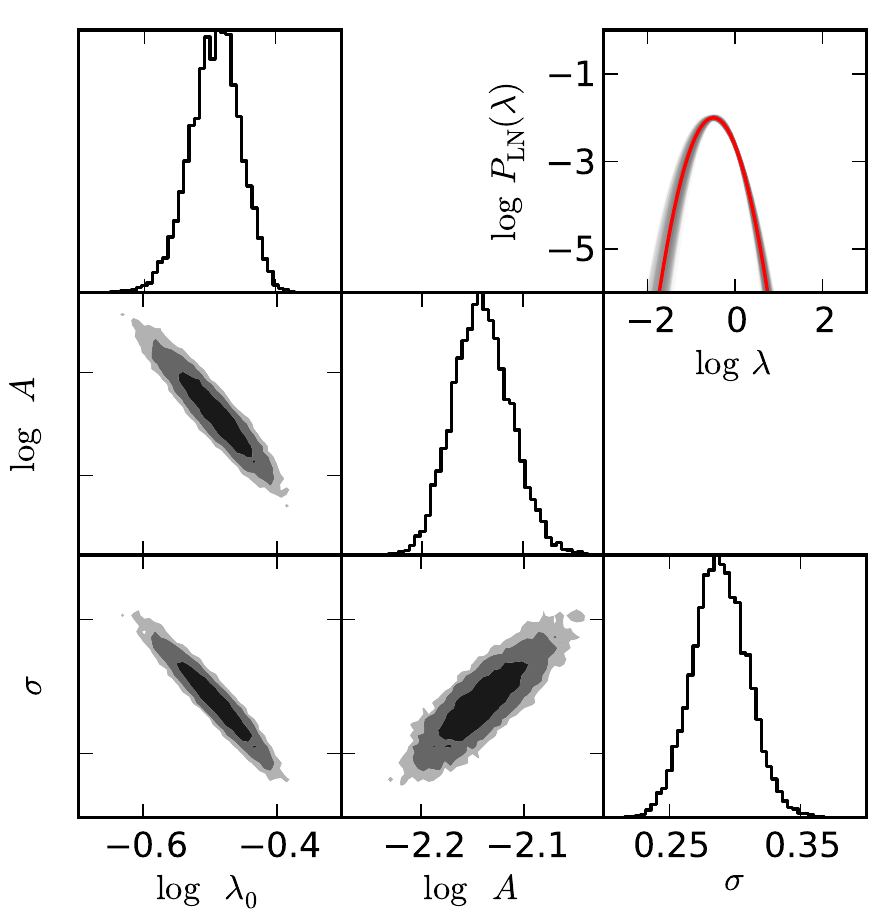}}
\end{center}
\caption{A representative corner plot for the Eddington model
  convolved with $P_{\rm LN}$ (the ``E+LN'' model) at \zfid. The other
  three model variations (and other redshifts) show qualitatively the
  same correlations among parameters, replacing $\sigma$ with $-\beta$
  where necessary. The red line shows the best-fit $P(\lambda)$
  distribution, with the shaded gray region generated from a random
  sample of points from the MCMC chain, plotted with some
  transparency. Parameter correlations are shown with 1, 2, and
  3-sigma regions (in dark, medium, and light gray).}
\label{fig:corner}
\end{figure}

\begin{figure}
\begin{center}
\resizebox{3.5in}{!}{\includegraphics{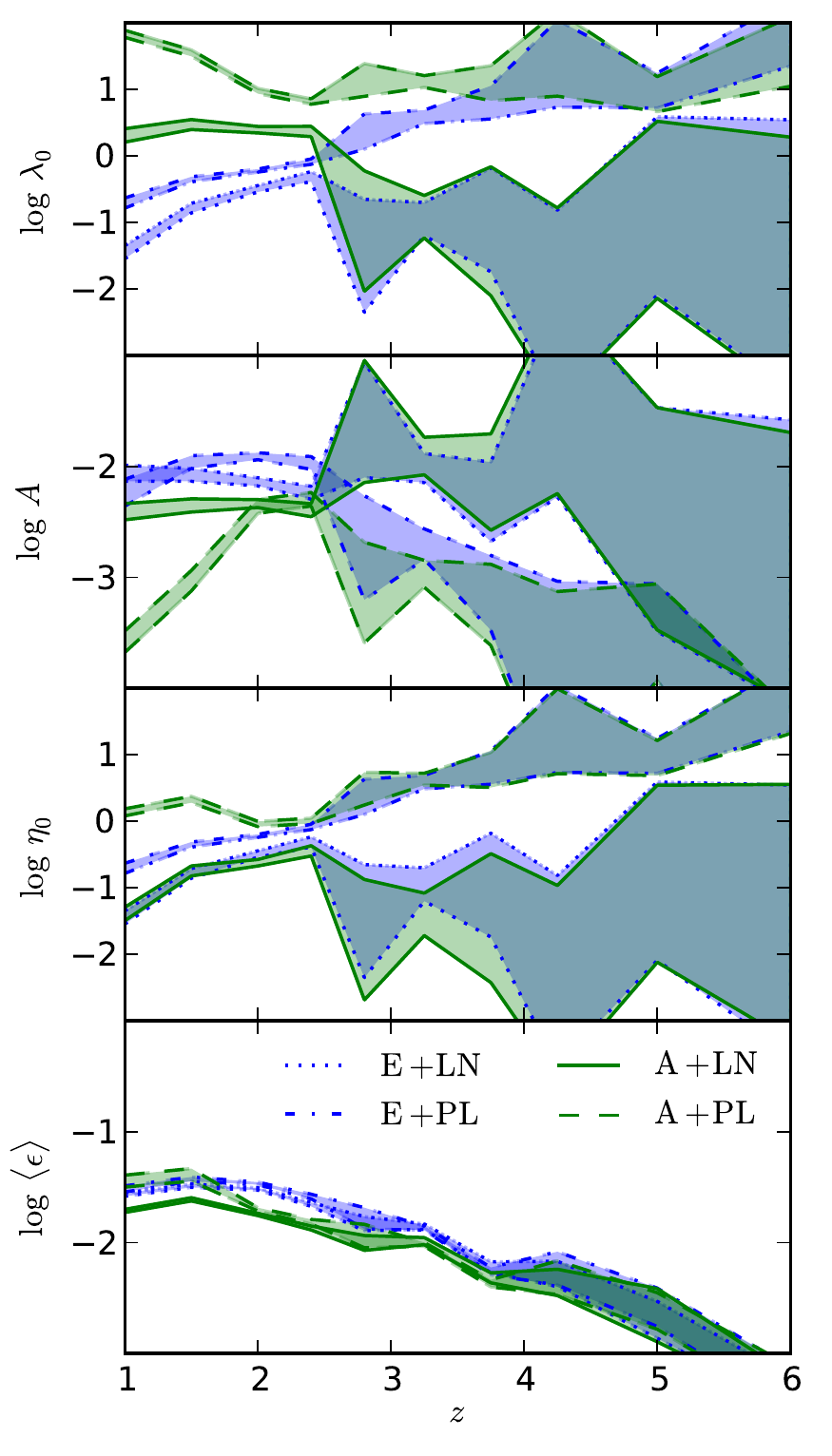}}
\end{center}
\caption{Model parameters vs redshift. The colors and line styles for
  each of the four model variations again follow the same conventions
  as in figures \ref{fig:model} and \ref{fig:multipanel}. The shaded
  regions represent the uncertainty in parameters from the MCMC fit.}
\label{fig:zparams}
\end{figure}

In figure \ref{fig:corner} we show an example of the MCMC fit results,
for the Eddington model using a $P_{\rm LN}$ distribution at \zfid. The
general shapes of the parameter correlations are the same for all
model variation and redshifts, including models using $P_{\rm PL}$ (when
$\sigma$ is replaced by $-\beta$). The best-fit parameters are highly
correlated, and we make particular note of the correlation between
$\lambda_0$ and $A$, which results in a very small error on
$\langle\lambda\rangle$. In a very rough approximation, we can write
the average $\lambda$ as
\begin{equation}
\log \, \langle\lambda\rangle \sim \log \, A - \log \, \lambda_0
\end{equation}
which follows the correlation of $A$ and $\lambda_0$ in the MCMC fit
results. This is notable because, as we see below, the efficiency
$\langle\epsilon\rangle$ is related to $\langle\lambda\rangle$. The
correlations of both $A$ and $\lambda_0$ with $\sigma$ are also fairly
strong, but do not directly impact any ``physical'' parameters of the
model. These correlations can be understood by referencing figure
\ref{fig:vary_params}: increasing $\sigma$ ``boosts'' the bright end
of the QLF, and leaves the faint end unchanged. To ``undo'' this
change and keep a good fit to the data, the QLF must be shifted up and
left, by increasing $A$ and decreasing $\lambda_0$.

Figure \ref{fig:zparams} shows the best-fit parameters $\lambda_0$ and
$A$ as a function of redshift for each model, along with the
characteristic Eddington ratio $\eta_0$ and average efficiency
$\langle\epsilon\rangle$. We can use the the specific growth rate
of the BH, $\psi_{\rm BH}\propto\Mbhdot/\Mbh$, to easily calculate $\eta_0$ and
$\langle\epsilon\rangle$ for all model variations:
\begin{align}
\eta_0 &= \frac{L_0}{L_{edd}} = \lambda_{edd} = \lambda_{acc}
\frac{L_{acc}}{L_{edd}} = \frac{\lambda_{acc}}{\psi_{\rm BH}}
\\ \langle\epsilon\rangle &= \frac{ \langle L \rangle}{L_{acc}}
= \langle \lambda_{acc} \rangle = \langle \lambda_{edd} \rangle
\psi_{\rm BH} \label{eq:eff}
\\ &\text{where} \; \psi_{\rm BH} \equiv \frac{L_{acc}}{L_{edd}} \propto
\frac{\Mbhdot}{\Mbh} \label{eq:psibh}
\end{align}
We have defined $\psi_{\rm BH}$ in units convenient to the problem at
hand, but it is analogous to the specific star formation rate of
galaxies. In reality, $\psi_{\rm BH}$ is a function of BH mass, but over
much of the mass range in question it is approximately constant,
similar to the specific star formation rate discussed in the
appendix. The efficiency we have defined here is closely related to
the radiative efficiency; however, radiative efficiency would
typically be defined in terms of the total mass inflow, as opposed to
the amount of mass ultimately accreted onto the black hole, which in
this case would be $\epsilon_{\rm rad}=L/(L+L_{acc})$. For small
values of $\epsilon$ the difference between these is small. When we
refer to ``the efficiency'', we are referring to the definition in
equation \ref{eq:eff} unless specified otherwise.

There are four notable features of figure \ref{fig:zparams}.
\begin{itemize}
\item The uncertainty of the fit parameters increases dramatically
  starting at redshift $z=2.8$.
\item The Eddington models (in blue) and accretion models (in green)
  give nearly identical predictions for the ``physical'' parameters,
  Eddington ratio and efficiency, while the resemblance is not as
  clear in $\lambda_0$ and $A$.
\item The uncertainty on the efficiency is much smaller than the
  uncertainty on the other parameters, and does not increase much with
  redshift.
\item The efficiency is nearly constant at around 3\% at $z=1$, but
  drops gradually with increasing redshift, while the Eddington ratio
  for the $P_{PL}$ models increases with redshift to greater than 10.
\end{itemize}

The second item, the similarity between Eddington and accretion
models, has already been discussed somewhat in \ref{sec:bright} and
\ref{sec:faint}. The ``intrinsic'' QLFs in the Eddington and accretion
models are very similar at low mass (or low growth rate), where our
approximation of constant $\psi_{\rm BH}$ is good. Although they are
somewhat different at high mass, this difference is ``washed out'' by
quasar variability when fitting the QLF. We will explore the validity
of the constant $\psi_{\rm BH}$ approximation further in section
\ref{sec:downsizing}, but find that it does not have much impact on
our analysis.

The first and third items are closely related, having to do with the
correlation between $\lambda_0$ and $A$. As discussed in
section \ref{sec:fit}, $\lambda_0$ can adjust the QLF left or right
while $A$ can adjust it up or down. Thus, if the observed QLF were a
single power-law at all scales, there would be a perfect degeneracy
between these two parameters. Instead, the QLF has a ``knee'', which
limits the degeneracy. However, that ``knee'' becomes less prominent
at high redshift, restoring some of the degeneracy. This is not
(necessarily) due to a failure to sample faint luminosities below the
``knee'' at high redshift, but is due to a combination of larger error
bars on the data and a less prominent ``knee'' inherited from the BHMF
and galaxy SMF.

An increased degeneracy between $\lambda_0$ and $A$ does not, however,
increase uncertainty in $\langle\lambda\rangle\sim\langle\epsilon\rangle$.
The efficiency is well constrained at all redshifts, given the
assumptions of our model. This illustrates the robustness of classic
arguments, such as in \citet{Soltan}, about the connection between
quasar luminosities and BH masses. This argument sets constraints on
the radiative efficiency of BHs by comparing the total integrated
luminosity of the QLF to the total integrated mass growth of the
BHMF. Since all of our fiducial model variations use the same BHMF,
output a similar QLF, and assume a single (approximately)
mass-independent efficiency, it is unsurprising that the efficiency is
very similar across model variations. (Variations beyond our fiducial
model, which consider different assumptions about the BHMF, would be
expected to give different values for the efficiency.) For any model,
correlations in the parameters of our fit may increase the uncertainty
of individual parameters such as $A$ and $\lambda_0$, but they do not
increase the uncertainty of the efficiency because it depends more
directly, in some sense, on the final shape of the QLF.

Finally, the fourth item raises several questions about the physical
implications of our model(s). Substantially super-Eddington accretion
may be physically questionable, and it can be argued that efficiency
and Eddington ratio may not be expected to evolve much with redshift
(or other parameters) if they are set largely by some universal
accretion physics.

The large Eddington ratios for the $P_{\rm PL}$ model variations may
be explained by the lack of ``adjustable'' scatter in those versions
of the model. If the bright end of the QLF is set by $P(\lambda)$, as
discussed in section \ref{sec:bright}, then $P_{\rm PL}$ has less
flexibility in fitting this portion of the QLF, as illustrated in
figure \ref{fig:vary_params}. Any additional source of scatter would
likely decrease the Eddington ratio. (Some examples of additional
sources of scatter include modifying $P_{\rm PL}$, adding scatter to
the BH-galaxy mass relationship, or accounting galaxy growth beyond
what is derived by our simple matching procedure, such as populations
of massive galaxies that are still rapidly growing.) This follows the
behavior shown in figure \ref{fig:corner} for $P_{\rm LN}$, where
increasing $\sigma$ correlates with decreasing $\lambda_0$ in the MCMC
fit.

The best-fit values of Eddington ratio and efficiency also depend on
our choice of fiducial model, the ``growth-based evolution'' model of
the BH-galaxy connection. The Eddington ratio is degenerate with
$\alpha_M$, while the efficiency is degenerate with $\alpha_G$. As we
discuss in section \ref{sec:ev} and in the appendix, our fiducial
model is only one possible choice. An increase in $\alpha_M$ at late
redshift could shift the best-fit Eddington ratio down, or a decrease
in $\alpha_G$ could shift the efficiency up. There is also the issue
of obscuration to consider, as the addition of a substantial
``obscured fraction'' to the calculation would increase the true
average efficiency from what is shown in figure \ref{fig:zparams}. We
refer to e.g. \citet{gil10} for a discussion of how quasar obscuration
may evolve with redshift, but note that it is unlikely for a simple
overall obscuration fraction (independent of mass and luminosity but
evolving with redshift) to account for nearly two orders of magnitude
in evolution of the efficiency. This may imply a ``true'' decrease in
efficiency at high redshift, but we do not make any strong conclusions
because of the potential for both obscuration and different choices of
$\alpha_G$ to modify our result.


\section{Beyond the fiducial model}
\label{sec:notfid}

\subsection{Choices of redshift evolution}
\label{sec:ev}

\begin{figure}
\begin{center}
\resizebox{3.5in}{!}{\includegraphics{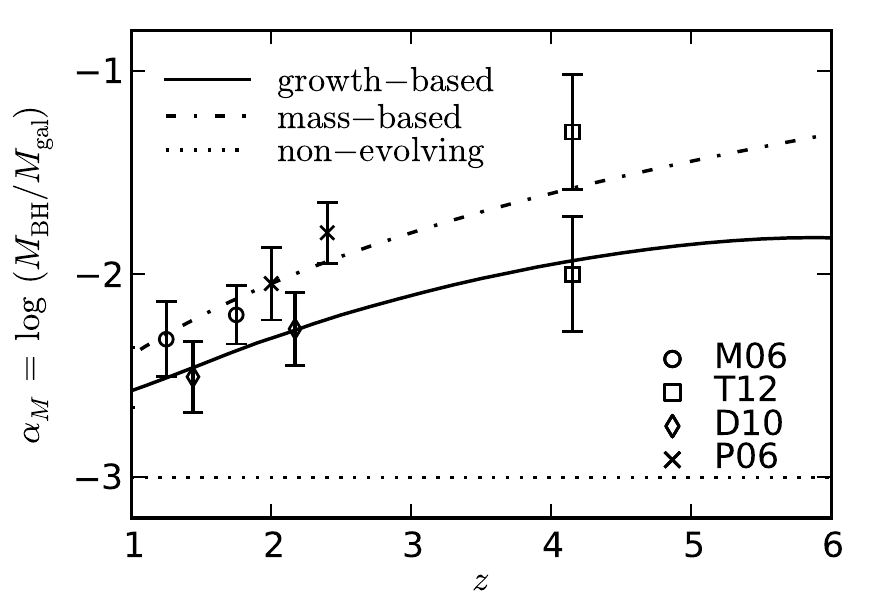}}
\end{center}
\caption{This plot shows $\alpha_M$ for each model of the BH-galaxy
  connection, as a function of redshift, compared to data on the
  $\Mbh/\Mgal$ relation. The data are a compilation from the following
  sources: \textbf{M06} (circle) is \citet{McL06}, \textbf{T12}
  (square) is \citet{Tar12} (with the two points representing two
  choices for estimating galaxy masses), \textbf{D10} (diamond) is
  \citet{Dec10}, \textbf{P06} (X) is \citet{Pen06}.}
\label{fig:alpha}
\end{figure}

\begin{figure}
\begin{center}
\resizebox{3.5in}{!}{\includegraphics{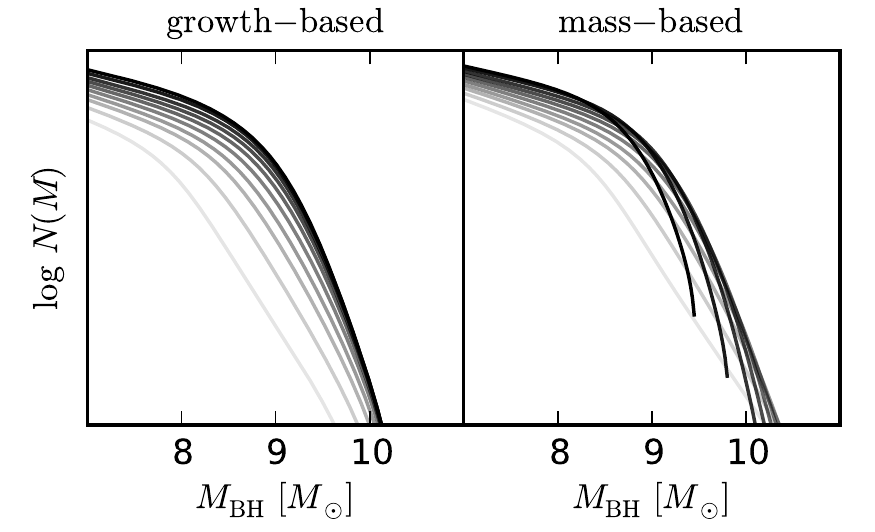}}
\end{center}
\caption{A comparison of the BHMF history for the growth-based
  evolution model (left panel) and mass-based evolution model (right
  panel), with the same conventions as the top left panel of figure
  \ref{fig:model}. The darkest lines represent $z=1$, and the lightest
  gray represent $z=6$. The right panel shows the inconsistency in the
  BHMF history for the mass-based evolution model, where the high mass
  end of the BHMF shrinks at low redshift.}
\label{fig:evolution}
\end{figure}

Throughout section \ref{sec:fid} we have been working within the
``growth-based evolution'' model of the BH-galaxy relationships across
redshift, but that is only one possible model choice. Now we would
like to explore the consequences of changing this model, first by
considering ``mass-based evolution'' and ``non-evolving'' models. We
will continue to enforce self-consistency in the BHMF history so that
we can make direct comparisons to all four fiducial model
variations.

We will also continue to neglect scatter in the BH-galaxy
relationships, so the differences in the BH properties of the three
models can be expressed entirely by the parameters $\alpha_M$ and
$\alpha_G$, which we defined in section \ref{sec:model_bh} as:
\begin{align}
\Mbh &= 10^{\alpha_M} \Mgal \nonumber
\\ \Mbhdot &= 10^{\alpha_G} \Mgaldot \nonumber
\end{align}
The appendix derives the behavior of $\alpha_M$ and $\alpha_G$ for the
three models, which we summarize here:
\begin{itemize}
\item Growth-based evolution
\begin{align}
\alpha_G(z) &= -3.5 + 2\log(1+z)
\\ \alpha_M(z,\Mgal) &\approx -3.2 + 2\log(1+z)
\end{align}
\item Mass-based evolution
\begin{align}
\alpha_M(z) &= -3.0 + 2\log(1+z)
\\ \alpha_G(z,\Mgal) &\approx -3.3 + 2\log(1+z)
\end{align}
\item Non-evolving
\begin{align}
\alpha_G = \alpha_M &= -3.0
\end{align}
\end{itemize}
The approximate mass-independent versions are more accurate at
low mass, and also ignore a slight additional redshift
evolution. Figure \ref{fig:alpha} shows $\alpha_M$ for each model
compared to a collection of $\Mbh/\Mgal$ data. Both evolving models
match the data fairly well, but the non-evolving model does not. We
also note the difference in our choices for the local values of
$\alpha$ in the two models: $\alpha_G=-3.5$ and $\alpha_M=-3.0$. These
are both motivated by observation (e.g. \citet{Che13} for $\alpha_G$;
e.g. \citet{HarRix04}, \citet{mcc13} for $\alpha_M$), but seem to have
some offset. In \citet{Che13}, several possible explanations are
mentioned for this offset, including substantial obscured accretion
activity and a need to account for bulge vs disk vs total galaxy
distinctions. However, in our model such an offset occurs naturally as
a consequence of evolution in the BH-galaxy relationship. (The value
of this offset is approximately 0.3 at $z=1$, but may be greater at
$z=0$. Also note that the precise local value of $\alpha_G=-3.5$ (or
$\alpha_M=-3.0$) may be freely tuned to better fit observations with
no impact on the overall model properties due to the degeneracy with
$\lambda_0$.)

Enforcing self-consistency in the BHMF also means that the
mass-based evolution model breaks down at low redshift. This was
the approach taken in CW13, and the inconsistent BH mass histories
were mentioned in that paper. Figure \ref{fig:evolution} compares the
BHMF history of the mass-based model with that of the growth-based
model. This figure uses the same convention for redshift as figure
\ref{fig:model}, with the darkest line corresponding to $z=1$ and the
lightest gray corresponding to $z=6$. Here it is easy to see the
``shrinking'' of the high-mass end of the BHMF in the mass-based
evolution model. This failure of self-consistency means that the
mass-based evolution model, when coupled with the accretion model for
determining the QLF, cannot fit the QLF at all; there is no
``intrinsic'' QLF to start from, because the accretion rate is
negative for nearly all BHs.

In all of our model variations (Eddington vs accretion models using
$P_{\rm LN}$ vs $P_{\rm PL}$), the parameter $\lambda_0$ is entirely degenerate
with either $\alpha_G$ (in the accretion models) or $\alpha_M$ (in the
Eddington models). With the exception of $\alpha_G$ in the mass-based
model, which is ruled out altogether for low redshift, we can
ignore any mass-dependence in $\alpha_M$ and $\alpha_G$ when
discussing this degeneracy. (We will justify this further in section
\ref{sec:downsizing}.) This means that for the purposes of fitting the
QLF, the only difference among successful models of the BH-galaxy
relationship evolution is a shift in the best-fit value of
$\lambda_0$, which depends only on redshift and compensates for the
shift in $\alpha_M$ or $\alpha_G$. For example, in the non-evolving
model, the best-fit values of Eddington ratio and efficiency would be
shifted up from their values in figure \ref{fig:zparams} by a factor
of approximately $2\log(1+z)$.

In summary, there are several potential concerns to consider in
choosing a model for the BH-galaxy relationship:
\begin{itemize}
\item Ensuring self-consistent $\alpha_M$ and $\alpha_G$.
\item Matching with observations of the $\Mbh/\Mgal$ and
  $\Mbhdot/\Mgaldot$ relations, at each redshift.
\item Avoiding substantially super-Eddington accretion or unrealistic
  efficiency.
\item Tuning the model to obtain Eddington ratios or efficiencies that
  do (or do not) evolve with redshift in some desired way.
\end{itemize}
Our fiducial model succeeds with the first three concerns listed,
while tuning the model to keep either Eddington ratio or efficiency
constant in redshift would require a more thorough exploration of the
``model space'' for $\alpha_G$ and $\alpha_M$ than our three example
choices. We note that keeping both Eddington ratio and efficiency
constant within this framework may not be possible, since $\alpha_M$
and $\alpha_G$ cannot be tuned independently.

While a general trend of increasing $\alpha$ at increasing redshift
is helpful both in matching the observed $\Mbh/\Mgal$ data and in
avoiding concerns about super-Eddington accretion and high efficiency,
it does present a potential challenge for BH seeding models. We will
not discuss this issue further, but it is an important one to consider
for connecting the model to redshifts beyond $z=6$.

\subsection{High mass objects and ``downsizing''}
\label{sec:downsizing}

We return to the growth-based evolution model to discuss the impact of
``downsizing'' on specific growth rates and related model
assumptions. At several points throughout the paper, we have used the
approximation that the specific growth rate of both galaxies ($\psi$
in the appendix) and black holes ($\psi_{\rm BH}$) is independent of
mass. This then allows us to assume that $\alpha_G$ and $\alpha_M$ are
both independent of mass as well, and results in the simple conversion
between Eddington ratio and efficiency in section \ref{sec:zparams}.

However, a mass-independent $\alpha_M$ and $\psi_{\rm BH}$ are unlikely to
be the case. In our model, $\psi_{\rm BH}$ is roughly independent of
mass at the low-mass end, but drops off at high mass, eventually
dropping sharply to zero for objects that are no longer growing at
all. The mass scale at which this occurs gets smaller at smaller
redshift, meaning the impact of this ``downsizing'' is largest at
small redshifts.

At high mass, where $\psi_{\rm BH}=0$, the parameter conversions from
section \ref{sec:zparams} give zero for the Eddington ratio in the
accretion models, and infinity for the efficiency in the Eddington
models. Infinite radiative efficiency is clearly not physically
reasonable, so we adjust the Eddington model to ``turn off'' these
high mass BHs. This is done by simply truncating the BHMF to include
only objects with a nonzero growth rate before we use it to derive the
``intrinsic'' QLF. The impact of this adjustment on the observed QLF
is completely negligible; the Eddington models fit the QLF equally
well, with the \emph{same} parameters, whether we truncate the BHMF or
not. If we go a step further and truncate the BHMF to exclude masses
where the growth rate is no longer increasing with mass (which
represents the mass scale at which the \emph{specific} growth rate
begins to drop rapidly towards zero), the effect on the QLF is still
negligible, and is only detectable at all at very low redshift and
very high luminosity.

Physically, the difference between the Eddington and accretion models
is most relevant in a limited range of masses, where we find objects
in ``transition'': their growth is slower than the characteristic
specific growth rate at that redshift, but has not stopped
completely. The Eddington model assumes these objects shine with the
same distribution of Eddington ratios as ``normal'' objects, which
would require a larger radiative efficiency. The accretion model
assumes these objects have the same distribution of $\lambda_{acc}$,
and hence the same average efficiency, as ``normal'' objects, which
results in a lower Eddington ratio. In principle, a hybrid model is
also possible, which holds both Eddington ratio and average efficiency
fixed, but instead adjusts the duty cycle (the normalization of the
$P(\lambda)$ distribution) for these ``transition'' objects to keep
everything self-consistent. It is virtually impossible, however, to
tell the difference between models at the level of the QLF, because of
the effects mentioned in section \ref{sec:bright}. Subtle effects such
as this at the high mass end are entirely ``washed out'' by the scatter
contained in the luminosity distributions $P(\lambda)$. (This scatter
is also what allows us to ``turn off'' the zero-growth BHs entirely in
the Eddington model without impacting the QLF.) In other words, for
the purposes of fitting the observed QLF the Eddington and accretion
models are essentially equivalent.

Another effect of a mass-dependent $\psi_{\rm BH}$ (and thus $\alpha_M$)
is curvature in the $\Mbh/\Mgal$ relationship. This effect is also
relevant only at high mass and low redshift. Even at $z\approx1$, we
find only a factor of 2 increase in $\Mbh$ at the largest mass scales,
compared to the linear $\Mbh/\Mgal$ relation. This is easily
consistent with the current uncertainty in observational measurements
of the local BHMF (e.g. \citet{mcc13}).

\subsection{Quasar host properties and BH-galaxy scatter}
\label{sec:hosts}

\begin{figure}
\begin{center}
\resizebox{3.5in}{!}{\includegraphics{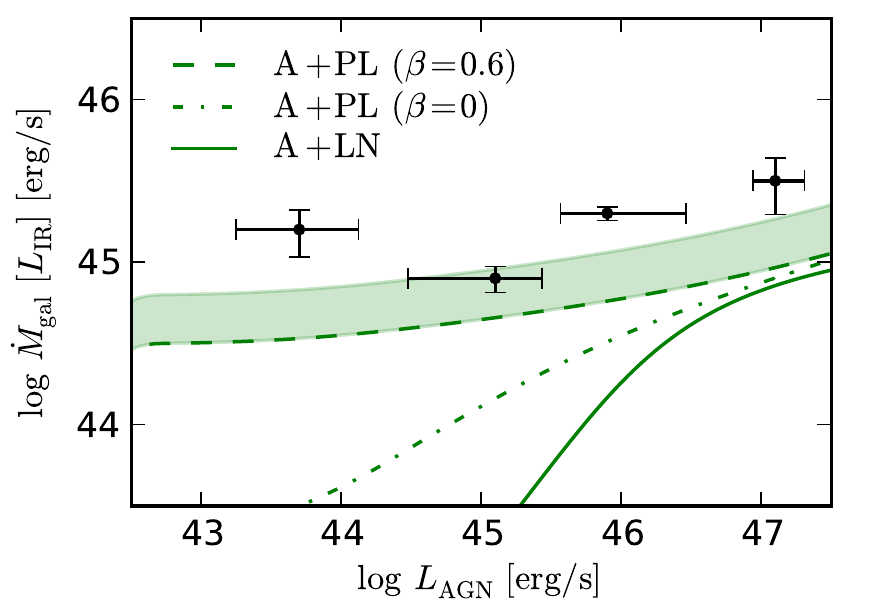}}
\end{center}
\caption{Average galaxy SFR (expressed in terms of  $L_{\rm IR}$) in
  bins of quasar luminosity. All lines shown are for the fiducial
  (growth-based evolution) accretion model at $z=2$. The solid line
  shows the model with $P_{\rm LN}$. The dash-dotted line shows the
  model with a flat slope in $P_{\rm PL}$. The dashed line shows the
  model with a steep slope in $P_{\rm PL}$. Because we use the total
  mass growth rate as an approximation to the SFR, neglecting stellar
  mass loss and other effects, we expect all of these lines to
  represent a low estimate of the true SFR. The dashed green line (our
  ``best-fit'' model in this case) is shown with a shaded region
  representing the expected correction (up to a factor of 2) to our
  calculation. The data points are from \citet{Rosa12}, and show the
  observed $L_{\rm IR}$ for the redshift bin $1.5<z<2.5$.}
\label{fig:gal_vs_lum}
\end{figure}

There are many observables beyond the QLF that can be used to
investigate models of quasar demographics. We refer to H14 for a
detailed discussion of the important distinction between measuring
average quasar luminosity in bins of host property versus measuring
average host property in bins of quasar luminosity. We have
constructed a similar model here: by design, average quasar
luminosity has a simple linear correlation with either galaxy mass or
growth rate. The inverse calculation, on the other hand, is very
sensitive to the various sources of scatter in the model.

Figure \ref{fig:gal_vs_lum} shows the average host galaxy ``SFR'' as a
function of quasar luminosity, for three variations of our fiducial
(growth-based evolution) accretion model. We write ``SFR'' in
quotation marks because we have only calculated an approximation to
the real galaxy SFR; in addition to the crudeness of our
constant-number-density matching procedure, we are neglecting factors
such as stellar mass loss. Our ``best-fit'' model in this case is the
variation with $P_{\rm PL}$ and a steep $\beta$: we show it with a
shaded band representing the region between our approximate SFR, which
is likely to be a low estimate, and the ``true'' SFR which may be
larger by a factor of 2 or more. (See e.g. \citet{Beh12}.) Our results
are similar to those in figure 3(b) of H14, except that we show a
weaker correlation at the high luminosity end. Although we show only
\zfid $\;$in figure \ref{fig:gal_vs_lum}, we do reproduce the trend in H14
that the ``characteristic'' SFR of low-luminosity quasars decreases
with decreasing redshift. However, the trend reverses above about
$z=3$ and begins to decrease with \emph{increasing} redshift, meaning
the characteristic SFR for low-luminosity quasars has a peak around
$2<z<3$.

Unlike the QLF, the measurement in figure \ref{fig:gal_vs_lum} is
quite sensitive to the difference between models using $P_{\rm LN}$ and
those using $P_{\rm PL}$, and to the slope $\beta$ of $P_{\rm PL}$. The weak
correlation of SFR and quasar luminosity (for low-luminosity quasars)
is good evidence for a $P(\lambda)$ that not only resembles $P_{\rm PL}$,
but has a fairly steep power-law $\beta$. The correlation at high
luminosity, on the other hand, is very similar for all of the models,
because of how well-constrained the bright end of $P(\lambda)$ is by
the QLF. However, $P(\lambda)$ itself is only well-constrained because
of our decision to neglect scatter in all of our other model
relationships. For measurements such as the one illustrated in figure
\ref{fig:gal_vs_lum}, the degeneracy between different sources of
scatter (between BH and galaxy properties, in $P(\lambda)$, between
galaxy mass and SFR, etc.) becomes important. For example, we could
imagine that there is substantial scatter between galaxy mass and SFR,
which would substantially boost the bright end of the ``intrinsic''
QLF in the accretion models. To compensate for this, and keep a good
fit to the observed QLF, we would have to decrease the amount of
``scatter'' in the bright end of $P(\lambda)$, by making the
exponential cutoff sharper somehow or decreasing $\sigma$ for
$P_{\rm LN}$. This would result in a substantially stronger correlation
at high luminosity in figure \ref{fig:gal_vs_lum}, because there would
be less scatter (at the bright end) between the SFR and QLF.

We can imagine our model, for the growth-based evolution and accretion
model case, as the following chain of calculations: halo mass
$\rightarrow$ galaxy mass $\rightarrow$ galaxy SFR $\rightarrow$
average BH accretion rate $\rightarrow$ BH ``intrinsic'' luminosity
$\rightarrow$ observed quasar luminosity. At each step there is
potentially some amount of scatter in the relations governing our
calculation, although we have neglected all sources of scatter outside
$P(\lambda)$ and the galaxy-halo mass relations. Since the total
amount of scatter (at the high mass/luminosity end, for a known halo
mass function) is well constrained by the QLF, we can only
``redistribute'' this scatter among the steps listed. If we imagine
making a measurement similar to figure \ref{fig:gal_vs_lum} for each
of the host properties, the correlation at high luminosity will depend
on the amount of scatter between that particular property and the end
of our ``chain'' of calculations. Thus, we expect the correlation
between host halo mass and quasar luminosity to be very weak
regardless of our other decisions about scatter. This is roughly
equivalent to saying that we do not expect the most luminous quasars
to live in halo environments very different from those of less
luminous quasars, which in turn impacts the luminosity dependence of
quasar bias and clustering (e.g. \citet{Tra12}, \citet{Fan13} as
mentioned in section \ref{sec:bright}). On the other hand, the
correlation between average BH accretion rate (or mass, if we were to
do this procedure for the Eddington models) and quasar luminosity
could be quite strong. This is roughly the situation described in
e.g. \citet{Hop09}, where very luminous quasars are typically objects
accreting at (or near) the Eddington luminosity. Again, this only
applies to the correlation for very luminous quasars; if the models
with $P_{\rm PL}$ and steep $\beta$ are correct, then the very weak
correlations at low luminosity would be expected for any host property
calculated as in figure \ref{fig:gal_vs_lum}.


\section{Summary}
\label{sec:summary}

We have constructed a self-consistent model of quasar demographics,
which links BH growth to galaxy growth across redshift and derives BH
masses by integrating this growth history. This model has four
variations wtihin the fiducial model: ``intrinsic'' quasar
luminosities are tied to either BH mass (the ``Eddington model'') or
BH growth rate (the ``accretion model''), and quasar variability is
modeled by either a ``scattered lightbulb'' model (a lognormal
luminosity distribution) or a ``luminosity-dependent lifetime'' model
(a power-law distribution with arbitrary lower bound and exponential
cutoff), shown schematically in figure \ref{fig:model}.

All four variations successfully fit the observed QLF (shown in figure
\ref{fig:multipanel}), despite being physically distinct models. The
``Eddington'' and ``accretion'' models are made nearly impossible to
distinguish by the similarity in their ``intrinsic'' QLFs at low
luminosity, which stems from the very weak dependence of specific
growth rates on mass (for both BHs and galaxies), and the fact that
the ``intrinsic'' QLF derived from the BHMF may be truncated to
include only objects with nonzero growth without impacting the fit to
the QLF. The remaining differences at high luminosity are washed out
by scatter, so that both models can fit the bright end of the QLF
despite differences in the bright ends of the ``intrinsic'' QLFs, and
regardless of whether the largest (non-growing) objects in the BHMF
are included.

The ``scattered lightbulb'' and ``luminosity-dependent lifetime''
models are difficult to distinguish because they are extremely similar
at the bright end of the luminosity distribution, which is relatively
well constrained by the QLF. Their main difference is in the faint end
of the distribution, which is very poorly constrained by the
QLF. However, measurements of the correlations between quasar
luminosity and host property (galaxy mass, SFR, BH mass, etc.) are
more sensitive to the details of various sources of scatter than the
QLF. Our model predicts, by design, a straightforward linear
correlation when measuring average quasar luminosity in bins of host
property. The inverse measurement, of average host property in bins of
quasar luminosity, is much more sensitive to scatter, with increased
scatter resulting in a weaker correlation. Weak correlations at low
luminosity are good evidence for ``luminosity-dependent lifetime''
models, with weaker correlations implying steeper power-law slopes. At
high luminosity, the strength of the correlation can potentially be
used to distinguish between different sources of scatter, such as
scatter in the BH-galaxy relationship (which we have neglected) vs
scatter in $P(\lambda)$.

Individual parameters in our model, particularly the characteristic
Eddington ratio and normalization of the luminosity distribution, have
increasing degeneracy at high redshift due to the ``softening'' of the
``knee'' of the QLF. The average efficiency, on the other hand,
is comparatively well-constrained at all redshifts and has very
similar best-fit values for all four model variations. While the
instantaneous efficiency may still be some strong function of
luminosity in our model, the average efficiency is well constrained by
the oberved QLF and BHMF alone.

Finally, requiring self-consistent redshift evolution in the BH-galaxy
relationship gives important constraints on the degeneracies in the
model. Based on fitting the QLF alone, there is a perfect degeneracy
between e.g. the normalization of $\Mbh/\Mgal$ ($\alpha_M$) and the
characteristic Eddington ratio ($\lambda_0$ in the Eddington
model). However, figure \ref{fig:alpha} illustrates the need for some
form of redshift evolution in $\alpha_M$ to match the observed
$\Mbh/\Mgal$ evolution. As discussed in the appendix, it is impossible
to keep a strictly linear relationship in both $\Mbh/\Mgal$ and
$\Mbhdot/\Mgaldot$ while including redshift evolution, so our fiducial
``growth-based'' model implies some curvature in $\Mbh/\Mgal$. A
``mass-based'' model, on the other hand, results in an inconsistent
BHMF history (illustrated in \ref{fig:evolution}), so we can rule it
out within our model framework. The curvature we predict in
$\Mbh/\Mgal$ is slight, restricted to high mass, and increases with
decreasing redshift.

A general conclusion of our model is that there are substantial
degeneracies within the ``model space'' of simple quasar demographics
models. We've explored three types of observation that are in some
sense ``orthogonal'' and helful to breaking these model degeneracies:
\begin{itemize}
\item The QLF can be fit equally well by many models, and is a
  well-studied output of large-scale redshift surveys, so it serves as
  an input to the model, fixing the best-fit model parameters (within
  certain degeneracies) which can then be used to ``predict'' other
  observables.
\item The $\Mbh/\Mgal$ relation allowed us to fix a ``fiducial model''
  of the BH-galaxy relationship, which would otherwise be quite
  unconstrained due to degeneracies between $\alpha_M$ (or $\alpha_G$)
  and $\lambda_0$. Making the connection across redshifts to require a
  self-consistent BHMF history further refines this model by ruling
  out our mass-based evolution in favor of growth-based evolution.
\item Measurements of average host property in bins of quasar
  luminosity are much more sensitive to the various sources of scatter
  in the model than the QLF. Weak correlations at the low luminosity
  end are evidence for models wherein quasars spend a large amount of
  time at relatively low luminosity, while comparing the correlations
  at high luminosity of different host properties may help identify
  which sources of scatter are most relevant to very luminous quasars.
\end{itemize}
The most persistent ``degeneracy'' in our model is between our
``Eddington'' and ``accretion'' models. To some degree, this
represents a true physical equivalence in our overall model, because
the BH mass and growth rate are roughly proportional to each other
over much of the relevant mass range. At very high mass, however,
``downsizing'' results in massive objects that are no longer growing,
which would imply infinite radiative efficiency in the simplest
version of the Eddington model. We showed that adjusting the Eddington
model to ``turn off'' those high mass objects has no impact on our model
predictions, except for small effects at our lowest redshifts. A more
sophisticated method of connecting BHs and galaxies across redshifts
could extend our model to lower redshift, where the difference between
Eddington and accretion models may no longer be negligible. 

To extend the model to $z<1$, following the methods of \citet{Beh12}
in more detail (e.g. by following halo merger trees instead of
matching galaxies at constant number density on the SMF) is one
possible way to obtain the necessary self-consistent galaxy histories
in that redshift range. More generally, any method that connects
galaxies across redshifts in a self-consistent way, tracking both mass
and growth rates, would be suitable for our model framework. Breaking
the various ``degeneracies'' in our model illustrates the value of a
diverse data set spanning multiple redshifts, luminosity ranges, and
measurable quantities, such as can be provided by large-scale redshift
surveys.


\section*{Acknowledgments} 

We would like to thank the referee Dr. Ryan Hickox for comments and
suggestions which improved the clarity of the paper. We also thank Tom
Targett for the compilation of data in Figure \ref{fig:alpha}, and
thank David Rosario for the data in Figure \ref{fig:gal_vs_lum}. This
work was supported by NASA. C.C. acknowledges support from Sloan and
Packard Foundation Fellowships. This work made extensive use of the
NASA Astrophysics Data System and of the {\tt astro-ph} preprint
archive at {\tt arXiv.org}.

\appendix
\section{Redshift evolution}

Our fiducial model, as well as the variations we consider in section
\ref{sec:notfid}, enforces self-consistent BHMF growth across
redshifts by associating objects at constant number density. We define
two parameters, $\alpha_M$ and $\alpha_G$, which encode the BH-galaxy
relationship, noting that in general they are not the same, and can
contain both redshift and mass dependence.
\begin{align}
\Mbh &= 10^{\alpha_M} \Mgal
\\ \Mbhdot &= 10^{\alpha_G} \Mgaldot
\\ \text{where } \alpha_M(z,\Mgal) &\ne \alpha_G(z,\Mgal)
\end{align}
The simplest model to consider is the ``non-evolving'' model, where
the $\Mbh/\Mgal$ relationship does not evolve with redshift and
is a simple linear relationship.
\begin{equation}
\alpha_M = \alpha_G = \alpha_0
\end{equation}
The non-evolving model we consider in the main text takes
$\alpha_0=-3.0$. (We will use $\alpha_0$ throughout the appendix to
note a constant value, with no mass or redshift dependence.)

Another choice is the model from CW13, which adds a simple redshift
evolution to $\alpha_M$. We call this the ``mass-based evolution''
model, and again use $\alpha_0=-3.0$ for the example in the main text
of the paper. The $(1+z)^2$ scaling, used here and in CW13, is chosen
as a possible broad match to observational data (as shown in figure
\ref{fig:alpha}).
\begin{align}
\Mbh &= 10^{\alpha_0} \Mgal (1+z)^2
\\ &= 10^{\alpha_M} \Mgal \label{eq:alphaM1}
\\ \implies \alpha_M &= \alpha_{0} + 2 \log (1+z) =
\alpha_M(z) \label{eq:alphaM2}
\end{align}
With this choice of redshift evolution for $\alpha_M$, we can then
derive $\alpha_G$ by taking the time derivative of equation
\ref{eq:alphaM1} :
\begin{align}
\Mbhdot &= 10^{\alpha_M} \Mgaldot \left( 1 + (\ln 10) \dot{\alpha}_M
\frac{\Mgal}{\Mgaldot} \right)
\\ &= 10^{\alpha_G} \Mgaldot
\\ \implies \alpha_G &= \alpha_M + \log \left( 1 + (\ln 10)
  \dot{\alpha}_M \frac{\Mgal}{\Mgaldot} \right)
\\ &= \alpha_G(z,\Mgal)
\end{align}
Here we can see that $\alpha_G$ must depend on both redshift and mass,
since the specific growth rate $\psi\equiv\Mgaldot/\Mgal$ depends on
mass. This mass dependence is stronger at high mass, so we can find an
approximation to $\alpha_G$ for small mass where $\psi$ is roughly
constant. In the following, we will use equation \ref{eq:alphaM2} to
evaluate $\dot{\alpha}_M$.
\begin{align}
\alpha_G &= \alpha_M + \log \left( 1 + (\ln 10)
  \frac{\dot{\alpha}_M}{\psi} \right)
\\ &= \alpha_M + \log \left( 1 + \frac{2}{1+z} \frac{\dot{z}}{\psi}
  \right)
\\ &= \alpha_M + \log \left( 1 - \frac{2H_0}{\psi}
  \sqrt{ \Omega_m (1+z)^{3} + \Omega_\Lambda} \right)
\\ &\approx \alpha_M -0.3
\end{align}
Where the $-0.3$ offset from $\alpha_M$ is a very rough approximation,
and neglects both the mass dependence and the additional redshift
dependence (beyond the redshift dependence of $\alpha_M(z)$). For our
analysis, we find $\alpha_G$ numerically by applying $\alpha_M$ to the
SMFs, then subtracting across redshifts.

Our fiducial model involves giving a simple redshift evolution to
$\alpha_G$, then integrating the masses across redshift to obtain
$\alpha_M$. We call this the ``growth-based evolution'' model. The
same general reasoning applies to this model as to the ``mass-based
evolution'' model, with $\alpha_G$ and $\alpha_M$ switching roles. The
end result is:
\begin{align}
\Mbhdot &= 10^{\alpha_0} \Mgaldot (1+z)^2 
\\ \implies \alpha_G &= \alpha_0 + 2\log(1+z) = \alpha_G(z)
\\ \implies \alpha_M &= \alpha_M(z,\Mgal) 
\\ &\approx \alpha_G(z) + 0.3
\end{align}
In the text, we use the growth-based evolution model with
$\alpha_0=-3.5$ as our fiducial model.

\end{document}